\documentclass[sigconf,  natbib=false]{acmart}

\usepackage{balance}

\def\BibTeX{{\rm B\kern-.05em{\sc i\kern-.025em b}\kern-.08emT\kern-.1667em\lower.7ex\hbox{E}\kern-.125emX}}
    
\copyrightyear{2019}
\acmYear{2019}
\setcopyright{acmlicensed}
\acmConference[CCS '19]{2019 ACM SIGSAC Conference on Computer and Communications Security}{November 11--15, 2019}{London, United Kingdom}
\acmBooktitle{2019 ACM SIGSAC Conference on Computer and Communications Security (CCS '19), November 11--15, 2019, London, United Kingdom}
\acmPrice{15.00}
\acmDOI{10.1145/3319535.3363225}
\acmISBN{978-1-4503-6747-9/19/11}

\usepackage[backend=biber, style=trad-plain]{biblatex}
\usepackage{algorithm, algpseudocode, amsmath, amssymb, booktabs, color, comment, enumitem, graphicx, hyperref, listings, multirow, siunitx, subcaption, tikz, xspace}

\graphicspath{{figure/}}

\newcommand{\fn}[1]{\mbox{\textsc{#1}}}

\newcommand{\file}[1]{\textit{#1}\xspace}

\newcommand{\name}{Matryoshka\xspace}

\begin{document}

\fancyhead{}


\title{\name: Fuzzing Deeply Nested Branches}

\author{Peng Chen}
\affiliation{%
  \institution{ByteDance AI Lab}
}
\email{spinpx@gmail.com}

\author{Jianzhong Liu}
\affiliation{%
  \institution{ShanghaiTech University}
}
\email{liujzh@shanghaitech.edu.cn}

\author{Hao Chen}
\affiliation{%
  \institution{University of California, Davis}
}
\email{chen@ucdavis.edu}

%

%
\begin{abstract}
\sloppy 
  Greybox fuzzing has made impressive progress in recent years, evolving from heuristics-based random mutation to solving individual branch constraints. However, they have difficulty solving path constraints that involve deeply nested conditional statements, which are common in image and video decoders, network packet analyzers, and checksum tools. We propose an approach for addressing this problem. First, we identify all the control flow-dependent conditional statements of the target conditional statement. Next, we select the taint flow-dependent conditional statements. Finally, we use three strategies to find an input that satisfies all conditional statements simultaneously. We implemented this approach in a tool called \name \footnote{\name dolls are the set of wooden dolls of decreasing size placed one inside another, which emblematizes the deeply nested conditional statements that our tool can fuzz.} and compared its effectiveness on 13 open source programs with other state-of-the-art fuzzers. \name achieved significantly higher cumulative line and branch coverage than AFL, QSYM, and Angora. We manually classified the crashes found by \name into 41 unique new bugs and obtained 12 CVEs. Our evaluation demonstrates the key technique contributing to \name's impressive performance: among the nesting constraints of a target conditional statement, \name collects only those that may cause the target unreachable, which greatly simplifies the path constraint that it has to solve.
\end{abstract}

%
%

\ccsdesc[500]{Security and privacy~Software security engineering}
\ccsdesc[500]{Software and its engineering~Software testing and debugging}

%
\keywords{fuzzing, optimization, taint analysis, vulnerability detection}

%
\maketitle

\section{Introduction}

Fuzzing is an automated software testing technique that has successfully found many bugs in real-world software.
Among various categories of fuzzing techniques, coverage-based greybox fuzzing is particularly popular, which prioritizes branch exploration to trigger bugs within hard-to-reach branches efficiently. Compared with symbolic execution, gray box fuzzing avoids expensive symbolic constraint solving and therefore can handle large, complex programs.

AFL~\cite{afl} is a rudimentary greybox fuzzer. It instruments the program to report whether the current input has explored new states at runtime.
If the current input triggers a new program state, then the fuzzer keeps the current input as a seed for further mutation~\cite{afl_technique}. 
However, since AFL mutates the input randomly using only crude heuristics, it is difficult to achieve high code coverage.

More recent fuzzers use program state to guide input mutation and showed impressive performance improvements over AFL, e.g.,  Vuzzer~\cite{vuzzer2017}, Steelix~\cite{li2017steelix}, QSYM~\cite{qsym}, and Angora~\cite{angora2018}. Take Angora for example. It uses dynamic taint tracking to determine which input bytes flow into the conditional statement guarding the target branch and then mutates only those relevant bytes instead of the entire input to reduce the search space drastically. Finally, it searches for a solution to the \emph{branch constraint} by gradient descent.

However, these fuzzers face difficulties when solving \emph{path constraints} that involve nested conditional statements. A branch constraint is the predicate in the conditional statement that guards the branch. The branch is reachable only if (1) the conditional statement is reachable, \emph{and} (2) the branch constraint is satisfied. A path constraint satisfies both these conditions. When a conditional statement $s$ is nested, $s$ is reachable only if some prior conditional statements $P$ on the execution path are reachable. If the branch constraints in $\{s\}\cup P$ share common input bytes, then while the fuzzer is mutating the input to satisfy the constraint in $s$, it may invalidate the constraints in $P$, thus making $s$ unreachable. This problem plagues the aforementioned fuzzers since they fail to track control flow and taint flow dependencies between conditional statements. Nested conditional statements are common in encoders and decoders for both images and videos, network packet parsers and checksum verifiers, which have a rich history of vulnerabilities.
Though concolic execution may solve some nested constraints, Yun et al.\ showed that concolic execution engines can exhibit over-constraining issues, which makes it too expensive to solve the constraints~\cite{qsym}, especially in real-world programs.

\autoref{fig:crc_example} shows such an example in the program \file{readpng}. 
The predicate on Line~\ref{lst:main:value_check} is nested inside the predicate on Line~\ref{lst:main:crc_check}.
\footnote{Although syntactically both Line~\ref{lst:main:crc_check} and Line~\ref{lst:main:value_check} are at the same level, Line~\ref{lst:main:value_check} is nested inside Line~\ref{lst:main:crc_check} in the control flow graph because the true branch of Line~\ref{lst:main:crc_check} is an immediate return.} 
It is difficult for the fuzzer to find an input that reaches the false branch of Line~\ref{lst:main:value_check} because the input has to satisfy the false branch of Line~\ref{lst:main:crc_check} as well. 
When a fuzzer tries to mutate the predicate on Line~\ref{lst:main:value_check}, it mutates only the input bytes flowing into \texttt{buffer[0]}, but this will almost surely cause the CRC check in \texttt{png\_crc\_finish()} to fail, which will cause Line~\ref{lst:main:crc_check} to take the true branch and return.

\begin{figure}[t]
\begin{lstlisting}[xleftmargin=0.5em]
// pngrutil.c, Line 2406
png_crc_read(png_ptr, buffer, length);
buffer[length] = 0;
if (png_crc_finish(png_ptr, 0) != 0) (*@\label{lst:main:crc_check}@*)
  return;
if (buffer[0] != 1 && buffer[0] != 2) { (*@\label{lst:main:value_check}@*) 
  png_chunk_benign_error(png_ptr, "invalid unit");
  return;
}
\end{lstlisting}
\caption{An example showing a nested conditional statement on Line \ref{lst:main:value_check}. It is difficult to find an input that reaches the false branch of Line~\ref{lst:main:value_check} due to the check on Line \ref{lst:main:crc_check}.}
\label{fig:crc_example}
\end{figure}

To evaluate whether current fuzzers have difficulty in solving path constraints involving nested conditional statements, we used Angora as a case study. We ran it on 13 open source programs, which read structured input and therefore likely have many nested conditional statements. \autoref{tbl:failedconstraints} shows that on all the programs, the majority of unsolved path constraints involve nested conditional statements. On five of the programs, more than 90\% of the unsolved constraints involve nested conditional statement.\footnote{Some conditional statements depend on other  conditional statements by control flow, but they do not share input bytes.} This suggests that solving these constraints will improve the fuzzer's coverage significantly.


We design and implement an approach that allows the fuzzer to explore deeply nested conditional statements. The following uses the program in \autoref{fig:reachability} as an example. Suppose the current input runs the false branch of Line~\ref{lst:main:br5}, and the fuzzer wishes to  explore the true branch of Line~\ref{lst:main:br5}.

\begin{figure}[t]
\begin{lstlisting}
void foo(unsigned x, unsigned y, unsigned z) {
  if (x < 2) {          (*@\label{lst:main:br1}@*)
    if (x + y < 3) {   (*@\label{lst:main:br2}@*)
      if (z == 1111) {  (*@\label{lst:main:br3}@*)
        if (y == 2222) { .... }   (*@\label{lst:main:br4}@*)
        if (y > 1) { .... }(*@\label{lst:main:br5}@*)
      }
    }
  }
}
\end{lstlisting}
\caption{A program demonstrating nested conditional statements. Line~\ref{lst:main:br5} depends on Line~\ref{lst:main:br1}, \ref{lst:main:br2}, and ~\ref{lst:main:br3} by control flow, and on Line~\ref{lst:main:br1} and \ref{lst:main:br2} by taint flow}.
\label{fig:reachability}
\end{figure}

\begin{enumerate}
\item \emph{Determine control flow dependency among conditional statements}.
The first task is to identify all the conditional statements before Line~\ref{lst:main:br5} on the trace that may make Line~\ref{lst:main:br5} unreachable. 
They include Line~\ref{lst:main:br1}, \ref{lst:main:br2}, and ~\ref{lst:main:br3}, because if any of them takes a different branch, then Line~\ref{lst:main:br5} will be unreachable. 
\autoref{sec:controlflow} will describe how we use intraprocedural and interprocedural post-dominator trees to find those conditional statements.

\item \emph{Determine taint flow dependency among conditional statements}. Among the conditional statements identified in the previous step, only those on Line~\ref{lst:main:br1} and \ref{lst:main:br2} have taint flow dependencies with Line~\ref{lst:main:br5}. This is because when we mutate $y$ on Line~\ref{lst:main:br5}, this may change the branch choice of Line~\ref{lst:main:br2} and hence making Line~\ref{lst:main:br5} unreachable. To avoid this problem, we must keep the branch choice of Line~\ref{lst:main:br2}, which may require us to mutate both $x$ and $y$, but this may change the branch choice of Line~\ref{lst:main:br1}. Therefore, Line~\ref{lst:main:br1} and \ref{lst:main:br2} have taint flow dependencies with Line~\ref{lst:main:br5}. By contrast, the branch choice of Line~\ref{lst:main:br3} never changes as we mutate $y$ to explore the true branch of Line~\ref{lst:main:br5}, so it has no taint flow dependencies with Line~\ref{lst:main:br5}. \autoref{sec:dataflow} will describe how we find those taint flow dependent conditional statements.

\item \emph{Solve constraints}. Finally, we need to mutate the input to satisfy several dependent conditional statements simultaneously. In other words, we need to find a new input that both reaches Line~\ref{lst:main:br5} and satisfies its true branch. We propose three strategies.

\begin{itemize}
\item The first strategy conservatively assumes that if we mutate any byte flowing into any conditional statements that Line~\ref{lst:main:br5} depends on, then Line~\ref{lst:main:br5} will become unreachable. So this strategy avoids mutating those bytes when fuzzing Line~\ref{lst:main:br5}.\footnote{This strategy fails to work on this example because the fuzzer is left with no input byte to mutate.} (\autoref{sec:prioritizereachability})

\item The second strategy artificially keeps the branch choices of all the conditional statements that Line~\ref{lst:main:br5} depends on when mutating the input bytes that flow into Line ~\ref{lst:main:br5}. When it finds a satisfying input, it verifies whether the program can reach Line~\ref{lst:main:br5} without altering branch choices. If so, then the fuzzer successfully solves this problem. Otherwise, the fuzzer will backtrack on the trace to try this strategy on Line~\ref{lst:main:br2} and Line~\ref{lst:main:br1}. (\autoref{sec:prioritizesatisfiability})

\item The last strategy tries to find a solution that satisfies all the dependent conditional statements. It defines a joint constraint that includes the constraint of each dependent conditional statement. When the fuzzer finds an input that satisfies the joint constraint, then the input is guaranteed to satisfy the constraints in all the dependent conditional statements. (\autoref{sec:joint})
\end{itemize}
\end{enumerate}

Our approach assumes no special structure or property about the program being fuzzed, such as magic bytes or checksum functions. Instead, our general approach to solving nested conditional statements can handle those special structures naturally.

We implemented our approach in a tool named \name and compared its effectiveness on 13 open source programs against other state-of-the-art fuzzers. 
\name found a total of 41 unique new bugs and obtained 12 CVEs in seven of those programs. \name's impressive performance is due not only to its ability to solve nested constraints but also to how it constructs these constraints. Traditional symbolic execution collects the predicates in all the conditional statements on the path. By contrast, \name collects the predicates in only those conditional statements that the target branch depends on by both control flow \emph{and} taint flow. Our evaluation shows that the latter accounts for only a small fraction of all the conditional statements on the path, which greatly simplifies the constraints that \name has to solve.

\section{Background}

Greybox fuzzing is a popular program testing method that incorporates program state monitoring with random input mutation to great effect. However, current state-of-the-art greybox fuzzers are unable to reliably and efficiently solve nested conditional statements.
Fuzzers using either heuristics (e.g., AFL) or principled mutation methods (e.g., Angora) do not have enough information about control flow and taint flow dependencies between conditional statements to devise an input that can satisfy all the relevant branch constraints. 
Other fuzzers utilizing hybrid concolic execution such as Driller experience performance hits due to concretizing the entire symbolic constraints of a path~\cite{qsym, zhao2019send}. QSYM is a practical concolic execution fuzzer, but it is tailored to solve only the last constraint on a path, thus facing the same challenge of solving nested conditional statements as Angora.

Using Angora as an example, we evaluated the impact of nested conditional statements on Angora's performance and analyzed the constraints in eight programs that Angora failed to solve in \autoref{tbl:failedconstraints}, where each constraint corresponds to a unique branch in the program. 
The second column shows what percentage of the unsolved constraints are nested, which depend on other conditional statements by control flow and taint flow (\autoref{sec:dataflow}). 
The third column shows what percentage of all the constraints, both solved and unsolved, are nested. 
\autoref{tbl:failedconstraints} shows that the majority of the unsolved constraints are nested, ranging from 57.95\% to nearly 100\%. 
It also shows that nested constraints account for a substantial portion of all the constraints, ranging from 44.14\% to 89.50\%. 
These results suggest that solving nested constraints could improve the coverage of greybox fuzzers substantially.

\begin{table}[t] 
    \caption{Percentage of nested constraints encountered by Angora}
    \begin{center}
    \begin{tabular}{lSS}
      \toprule
     \multirow{2}{*}{Program}  & \multicolumn{2}{c}{Percentage of nested constraints in}\\
    \cmidrule(lr){2-3}
    & {all unsolved constraints} & {all constraints} \\
    \midrule
    \file{djpeg}   & \SI{90.00}{\percent} & \SI{75.65}{\percent} \\
    \file{file}    & \SI{86.49}{\percent} & \SI{44.14}{\percent} \\
    \file{jhead}   & \SI{57.95}{\percent} & \SI{51.53}{\percent} \\
    \file{mutool}  & \SI{80.88}{\percent} & \SI{58.63}{\percent} \\
    \file{nm}      & \SI{84.32}{\percent} & \SI{68.16}{\percent} \\
    \file{objdump} & \SI{90.54}{\percent} & \SI{73.95}{\percent} \\
    \file{readelf} & \SI{84.12}{\percent} & \SI{70.50}{\percent} \\
    \file{readpng} & \SI{94.02}{\percent} & \SI{89.50}{\percent} \\
    \file{size}    & \SI{87.86}{\percent} & \SI{71.46}{\percent} \\
    \file{tcpdump} & \SI{96.15}{\percent} & \SI{78.98}{\percent} \\
    \file{tiff2ps} & \SI{75.56}{\percent} & \SI{62.18}{\percent} \\
    \file{xmlint}  & \SI{78.18}{\percent} & \SI{72.37}{\percent} \\
    \file{xmlwf}   & \SI{96.18}{\percent} & \SI{68.16}{\percent} \\
    \bottomrule
    \end{tabular}
    \end{center}
    \label{tbl:failedconstraints}
    \end{table}

\section{Design}
\label{sec:design}

\subsection{Problem}
\label{sec:problem}

State-of-the-art coverage-guided fuzzers, e.g., Angora~\cite{angora2018}, QSYM~\cite{qsym}, VUzzer~\cite{vuzzer2017} and REDQUEEN~\cite{redqueen}, explore new branches by solving branch constraints, where a branch constraint is the predicate in the conditional statement that guards the branch. This typically involve the following steps. First, identify the input bytes that affect each conditional statement using dynamic taint analysis or similar techniques. 
Then, determine how the input bytes should be mutated, such as calculating the gradient of the predicate and using gradient descent, matching magic bytes or resorting to using a symbolic execution solver.
Finally, execute the program with the mutated input and verify if this triggers the other branch in the conditional statement.

Although this approach is effective in solving many branch constraints, it fails when the target conditional statement becomes unreachable during input mutation.

\autoref{fig:reachability} shows an example. Let the variables \texttt{x}, \texttt{y}, and \texttt{z} contain different input bytes. 
Assume that the current input executes the false branch of Line~\ref{lst:main:br5}, and the goal is to explore the true branch of Line~\ref{lst:main:br5}. 
Then, the fuzzer determines, by dynamic byte-level taint analysis, that it needs to change the bytes in $y$. 
Consider two different initial values of $x$ and $y$.

\begin{enumerate}
\item $x=0$ and $y=1$. 
If the fuzzer mutates $y$ to 3, then the program will no longer reach Line~\ref{lst:main:br5} because Line~\ref{lst:main:br2} will take a different (false) branch. 
This renders the fuzzer helpless when solving the branch predicate, even though a satisfying assignment $y=2$ exists.

\item $x=1$ and $y=1$. 
In this case, no value of $y$ can satisfy the true branches of Line~\ref{lst:main:br1}, Line~\ref{lst:main:br2}, and Line~\ref{lst:main:br5} simultaneously, unless we also mutate $x$. 
However, since $x$ does not flow into the conditional statement on Line~\ref{lst:main:br5}, the fuzzer does not know that it should mutate $x$, so it can never find a satisfying assignment to explore the true branch of Line~\ref{lst:main:br5}, regardless of the algorithm used to solve the constraint.
\end{enumerate}

This example shows that to execute an unexplored branch, it is sometimes inadequate to mutate only the input bytes that flow into the conditional statement because doing so might render this statement unreachable. 
One could naively mutate all the input bytes, but that would increase the search space by many magnitudes to make this approach too expensive to be practical.

\subsection{Solution overview}

To overcome the problem in \autoref{sec:problem}, our key insight is that when we fuzz a conditional statement, we must find an input that both satisfies the branch constraint and keeps the statement reachable. Most fuzzers that explore branches by solving branch constraints consider only the satisfiability criterion but fail
to consider the reachability criterion. We propose the following steps to satisfy both criteria while mutating the input. Let $s$ be a conditional statement on the trace of the program on this input. Our goal is to mutate the input to let $s$ take a different branch. We call $s$ the \emph{target conditional statement} and say that the new input \emph{satisfies $s$}.

\begin{enumerate}
\item \emph{Determine control flow dependencies among conditional statements}. 
Identify all the conditional statements before $s$ on the trace that may make $s$ unreachable. 
For example, if $s$ is on Line~\ref{lst:main:br5} in \autoref{fig:reachability}, then if any of the conditional statements on Line~\ref{lst:main:br1}, \ref{lst:main:br2}, and ~\ref{lst:main:br3} takes a different branch, then Line~\ref{lst:main:br5} will be unreachable. 
We call these the \emph{prior conditional statements} of $s$, which $s$ depends by control flow. 
By contrast, no matter which branch Line~\ref{lst:main:br4} takes, Line~\ref{lst:main:br5} will always be reachable. 
\autoref{sec:controlflow} will describe this step in detail.

\item \emph{Determine taint flow dependency among conditional statements}. 
Among the prior conditional statements of $s$, identify those whose corresponding input bytes may have to be mutated to satisfy $s$. 
For example, let $s$ be Line~\ref{lst:main:br5} in \autoref{fig:reachability}. 
Among its three prior conditional statements, only those on Line~\ref{lst:main:br1} and \ref{lst:main:br2} contain bytes ($x$ and $y$) that may have to be mutated to satisfy $s$. 
We call these \emph{effective prior conditional statements}, which $s$ depends on by taint flow. 
By contrast, Line~\ref{lst:main:br3} contains no input bytes that may have to be mutated to satisfy $s$. 
\autoref{sec:dataflow} will describe this step in detail.

\item \emph{Solve constraints}. Mutate the bytes in the effective prior conditional statements to satisfy $s$. \autoref{sec:optimization} will describe this step in detail.
\end{enumerate}

\begin{figure*}[t]
  \centering
    \includegraphics[width=0.9\textwidth]{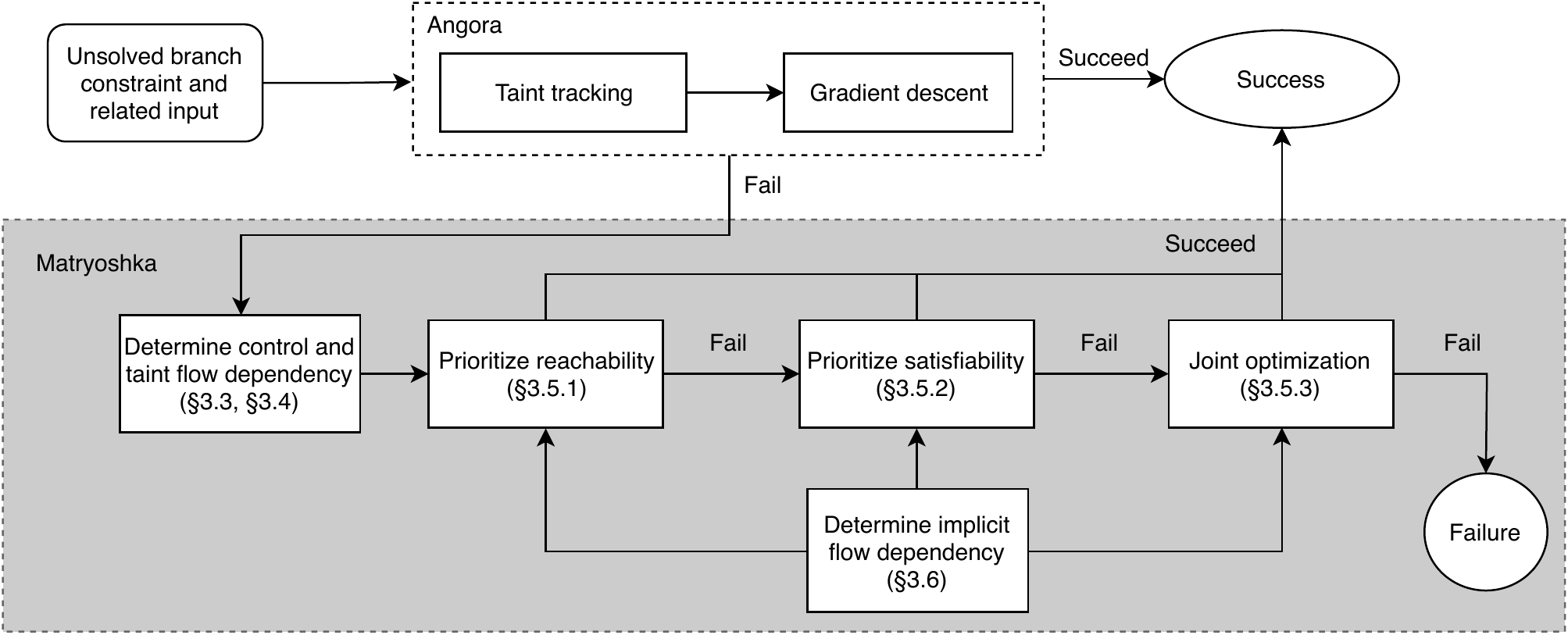}
    \caption{Overview of \name. In the figure, Angora represents any fuzzer capable of identifying constraints. When the fuzzer fails to solve a branch constraint guarding a new branch, \name determines whether the conditional statement is nested. If so, \name tries three optimization strategies: prioritizing reachability, prioritizing satisfiability, and joint optimization, during which it also identifies implicit flow dependencies when necessary.}
  \label{fig:flowchart}
  \end{figure*}

\autoref{fig:flowchart} shows an overall design of \name regarding how the strategies are used in the fuzzing process.

\subsection{Determine control flow dependency among conditional statements}
\label{sec:controlflow}

For each conditional statement $s$, we wish to identify all its \emph{prior conditional statements}, which are the conditional statements that, if taking a different branch, may cause $s$ to be unreachable. 
Let the \emph{immediate prior conditional statement} of $s$ on a trace be the last prior conditional statement of $s$, i.e., there is no prior conditional statement of $s$ between $r$ and $s$. 
Note that if $s$ is a prior conditional statement of $t$, and $t$ is of $u$, then $s$ is a prior conditional statement of $u$. 
This allows us to find all the prior conditional statements of $s$ transitively: starting from $s$, we repeatedly find the immediate prior conditional statement, and then take the union of all such statements.

We propose two different methods for finding the immediate prior statement that is in the same function and that is in a different function, respectively.
In our implementation for optimization, we cached all the found dependencies to avoid repeated computation.

\subsubsection{Intraprocedural immediate prior conditional statement}

\label{sec:intraprocedural}
Starting from a conditional statement $s$, we walk back on the trace. When we find the first conditional statement $r$
\begin{itemize}
\item that is in the same function, and
\item that $s$ does \emph{not} post-dominate~\cite{allen1970control} 
\end{itemize}
then $r$ is the immediate prior statement of $s$. Our implementation used the post-dominator trees produced by LLVM~\cite{llvm}.

If we cannot find such $r$, then $s$ has no intraprocedural immediate prior conditional statement, and we will search for its interprocedural immediate prior conditional statement, to be described in \autoref{sec:interprocedural}.

\subsubsection{Interprocedural immediate prior conditional statement}
\label{sec:interprocedural}
It would be straightforward to use interprocedural post-dominator trees for efficient handling, but unfortunately, LLVM does not provide such information, so we designed the following method for finding the interprocedural immediate prior conditional statement of $s$. 
Starting from $s$, we walk back on the trace to find the first conditional statement $r$ that satisfies all the following:

\begin{enumerate}
\item $r$ is in a different function (let us call it $f_r$) than $s$, and

\item $f_r$ is still on the stack (i.e., it hasn't returned) when $s$ is executing, and
  
\item Let $r_c$ be the last call instruction that $f_r$ executed. $r_c$ must exist because $r$ is in a deeper stack frame than $s$. 
If $r_c$ does not post-dominate $r$ (note that $r$ and $r_c$ are in the same function), then $r$ is the interprocedural immediate prior statement of $s$.
\end{enumerate}
\label{thm:interprocedural}

\subsubsection{Irregular interprocedural control flow}
\label{sec:irregular_flow}
Apart from function calls, the program could also exhibit irregular interprocedural control flows, for instance those involving \fn{exit} and \fn{longjmp} instructions. If a conditional statement $r$ has at least one branch that leads to a basic block that contains irregular flows, then we consider it to be the prior conditional statement of all the statements after itself even when its frame is not on the stack. If $s$ is a conditional statement after $r$, we add $r$ and $r$'s prior conditional statements to the set of $s$'s prior conditional statements. In LLVM, the basic blocks containing irregular interprocedural control flows are terminated with \fn{unreachable} instructions. 

\subsection{Determine taint flow dependency among conditional statements}
\label{sec:dataflow}

For each conditional statement $s$, \autoref{sec:controlflow} finds all its prior conditional statements $p(s)$. 
Let $b(s)$ be the set of input bytes that flow into $s$ where $s$ is one or more conditional statements. 
When we mutate the input, as long as no conditional statement in $p(s)$ takes a different branch, $s$ is guaranteed to be reachable. 
This seems to suggest that we should avoid mutating any byte in $b(p(s))$.

On the other hand, avoid mutating every byte in $b(p(s))$ may prevent the fuzzer from finding a satisfying assignment for $s$, as discussed in \autoref{sec:problem}. 
Take \autoref{fig:reachability} as an example. Let $s$ be Line~\ref{lst:main:br5}. By \autoref{sec:intraprocedural}, we determine that $p(s)$ consists of Lines~\ref{lst:main:br1}, \ref{lst:main:br2}, and \ref{lst:main:br3}. 
Therefore, $b(p(s))=\{x, y, z\}$. If we keep all the bytes in $b(p(s))$ immutable, then we are left with no input byte to mutate when trying to find an input to satisfy $s$.

The problem arises because \autoref{sec:controlflow} considers only control flow dependency among conditional statements, but it fails to consider whether taint flow dependencies exist between the conditional statements. 
We define the \emph{effective prior conditional statements} of $s$, $e(s)$, to be a subset of the prior conditional statements of $s$, where to find an input to satisfy $s$, we may have to mutate some bytes flowing into a statement in $e(s)$. 
In other words, if a prior conditional statement $r$ of $s$ is not also an effective prior conditional statement of $s$, then no byte flowing into $r$ needs to be mutated to satisfy $s$. 
This means that we may consider only the effective prior conditional statements and ignore the non-effective prior conditional statements.

Algorithm~\ref{alg:dataflow} shows the algorithm for computing effective prior conditional statements, which relies on the following property: if $r$ is an effective prior conditional statement of $s$, and $q$ is a prior conditional statement of $s$, and $q$ and $r$ share common input bytes, then $q$ is also an effective prior conditional statement of $s$.

\begin{algorithm}
\begin{algorithmic}[1]
\Function{FindEffectivePriorCondStmt}{$s, stmts$} 
\Comment{$s$: conditional statement being fuzzed; $stmts$: prior conditional statements of $s$. Returns: effective prior conditional statements of $s$}

\State Initialize a union-find data structure.
\ForAll{$stmt \in stmts$}
   \State $T \gets$ input bytes flowing into $stmt$.
   \State \fn{Union} all $t\in T$.
\EndFor

\State $O \gets \emptyset$
\State $b_s \gets$ any one byte flowing into $s$
\ForAll{$stmt \in stmts$}
    \State $b \gets$ any one byte flowing into $stmt$
    \If{\fn{Find}($b_s$) == \fn{Find}($b$)}
      \State $O \gets O \cup \{stmt\}$
    \EndIf
\EndFor

\State \Return $O$
\EndFunction
\end{algorithmic}
\caption{Find effective prior conditional statements}
\label{alg:dataflow}
\end{algorithm}

\subsection{Solve constraints}
\label{sec:optimization}

Section~\ref{sec:dataflow} determines the effective prior conditional statements for each conditional statement $s$. On the one hand, if we freely mutate the bytes flowing into any of them, $s$ may become unreachable. But on the other hand, we may be required to mutate some of those bytes to satisfy the unexplored branch of $s$. Therefore we need to determine which of those statements whose relevant input bytes we may mutate. We propose the following three alternative strategies. \name tries them in this order and imposes a time budget for each strategy to ensure overall efficiency.

\begin{enumerate}
\item Prioritize reachability
\item Prioritize satisfiability
\item Joint optimization for both reachability and satisfiability
\end{enumerate}

Each strategy identifies a constraint over a set of input bytes. Then, it uses gradient-based optimization to solve the constraint. These strategies provide benefits only when $s$ is nested, i.e., $s$ has  effective prior conditional statements. If $s$ is not nested, \name simply uses existing strategies from Angora or other fuzzers to solve the branch constraint. Therefore, \name exhibits better performance in solving nested conditional statements while having the same ability as other fuzzers to solve non-nested conditional statements.

\subsubsection{Prioritize reachability}
\label{sec:prioritizereachability}

This strategy pessimistically assumes that if we mutate any byte that flows into any effective prior conditional statements of a conditional statement $s$, $s$ may become unreachable. 
Therefore, this strategy ensures that $s$ is always reachable by avoiding mutating any byte that flows into any of $s$'s effective prior conditional statements. 
Formally, let $b(s)$ be the bytes flowing into $s$, and $b(e(s))$ be the bytes flowing into $s$'s effective prior conditional statements. 
Angora mutates all the bytes in $b(s)$, which may cause $s$ to become unreachable. 
By contrast, this strategy of \name mutates only the bytes in $b(s)\setminus b(e(s))$, i.e. all the bytes that flow into $s$ but that do not flow into any effective prior statement of $s$. 

Take the program in \autoref{fig:reachability} for example. 
When we fuzz $s$ on Line~\ref{lst:main:br2}, its only effective conditional statement is $t$ on Line~\ref{lst:main:br1}. 
$b(s)=\{x, y\}$. 
$b(e(s))=\{x\}$. 
Using this strategy, the fuzzer mutates only the bytes in $b(s)\setminus b(e(s))=\{y\}$.

However, this strategy fails when we fuzz $s$ on Line~\ref{lst:main:br5}. In this case, its effective prior statements consist of the statements on Line~\ref{lst:main:br1} and \ref{lst:main:br2}, so $b(s)=\{y\}$, $b(e(s))=\{x, y\}$, but $b(s)\setminus b(dp(s))=\emptyset$. Using this strategy, \name will fail to fuzz $s$ because it finds no byte to mutate.

\subsubsection{Prioritize satisfiability}
\label{sec:prioritizesatisfiability}

\begin{algorithm}
\begin{algorithmic}[1]
\Function{FindInput}{$s, stmts$} 
\Comment{$s$: the target conditional statement. $stmts$: effective prior conditional statements of $s$.}
\State
\Comment{Forward phase}
\State While keeping the branch choice of each $r\in stmts$, find an input $i$ that satisfies the target branch of $s$.
\State Run the program on $i$.
\If{$s$'s target branch is reachable}
\State \Return Success
\EndIf
\State
\Comment{Backtracking phase}
\State $B_I \gets \emptyset$
\Comment{Input bytes not to be mutated.}
\For{$stmt \in stmts$ in the reverse order on the trace}
\State $B \gets$ input bytes flowing into $stmt$
\State $B_2 \gets B \setminus B_I$
\State While keeping the branch choice of all $r\in stmts$ where $r$ appears before $stmt$ on the trace, find an input $i$ that satisfies the target branch of $stmt$, during which only the input bytes in $B_2$ may be mutated.
\State Run the program on $i$.
\If{$stmt$'s target branch is reachable}
\State \Return Success
\EndIf
\State $B_I \gets B_I \cup B$    
\EndFor

\State \Return Failure
\EndFunction
\end{algorithmic}
\caption{Find a satisfying input while prioritizing satisfiability}
\label{alg:backward}
\end{algorithm}

This strategy optimistically hopes that a mutated input that satisfies a conditional statement $s$ can also reach $s$. 
It has a forward phase followed by a backtrack phase. 
During the forward phase, it mutates the bytes flowing into $s$ while artificially keeping the branch choices of all the effective prior conditional statements of $s$, thereby guaranteeing that $s$ is always reachable. 
If it finds an input that satisfies the target branch of $s$, it runs the program on that input normally (without artificially fixing branch choices). 
If this trace still reaches $s$ and chooses the target branch, it succeeds. 
Otherwise, it enters the backtrack phase. 
During this phase, it starts from $s$ and then goes backward to fuzz each of the effective prior statements of $s$ in that order. 
When it fuzzes one such statement $r$, it avoids mutating any byte that may flow into $s$ or any effective prior conditional statement of $s$ that is after $r$. 
The process succeeds when the fuzzer successfully fuzzes all of these effective prior conditional statements. 
\autoref{alg:backward} shows this algorithm.

Take the program in \autoref{fig:reachability} as an example. 
When we fuzz $s$ on Line~\ref{lst:main:br5}, its effective prior conditional statements are on Line~\ref{lst:main:br2} and \ref{lst:main:br1}. 
Let the current input be $x=1, y=1$. 
Under this input, both Line~\ref{lst:main:br1} and \ref{lst:main:br2} take the true branch, and Line~\ref{lst:main:br5} takes the false branch. 
Our goal is to take the true branch on Line~\ref{lst:main:br5}. 
Using this strategy, during the forward phase, the fuzzer mutates $y$ while artificially forcing the program to take the true branch on both Line~\ref{lst:main:br1} and \ref{lst:main:br2}. 
If the fuzzer finds an assignment $y=2$ to satisfy the true branch of Line~\ref{lst:main:br5}, but since $x=1, y=2$ does not satisfy Line~\ref{lst:main:br2}, it enters the backtracking phase. 
During this phase, it will first fuzz Line~\ref{lst:main:br2}. 
Although this line is affected by two values $\{x, y\}$, since $y$ flows into Line~\ref{lst:main:br5}, the fuzzer will mutate $x$ only. 
If it finds a satisfying assignment $x=0$, it tries to run the program with $x=0, y=2$ without artificially forcing branch choices. 
Since this input reaches Line~\ref{lst:main:br2} and satisfies the target (true) branch, fuzzing succeeds.

By contrast, let us assume that the fuzzer finds a satisfying assignment $y=3$ when fuzzing Line~\ref{lst:main:br5}. 
During the backtrack phase, when fuzzing Line~\ref{lst:main:br2}, since it can mutate only $x$, it cannot find a satisfying assignment. Therefore, fuzzing of $s$ fails.

\subsubsection{Joint optimization for both reachability and satisfiability}
\label{sec:joint}

Both strategies in \autoref{sec:prioritizereachability} and \autoref{sec:prioritizesatisfiability} search for a solution to one constraint at a time. 
\autoref{sec:prioritizereachability} mutates only the input bytes that will not make the target conditional statement unreachable, while \autoref{sec:prioritizesatisfiability} tries to satisfy the target conditional statement and its effective prior conditional statements one at a time. 
However, they fail to find a solution where we must jointly optimize multiple constraints.

Let $s$ be the target conditional statement. 
Let $f_i(\boldsymbol{x}) \le 0, \forall i\in [1, n]$ represent the constraints of the effective prior conditional statements of $s$, and $f_o(\boldsymbol{x}) \le 0$ represent the constraint of $s$. $\boldsymbol{x}$ is a vector representing the input bytes. 
\autoref{tbl:trans} shows how to transform each type of comparison to $f\le 0$. 
Our goal is to find an $\boldsymbol{x}$ that satisfies all $f_i(\boldsymbol{x}) \le 0, i\in [0, n]$. 
Note that each $f_i(\boldsymbol{x})$ is a blackbox function representing the computation on the input $\boldsymbol{x}$ by the expression in the conditional statement $i$. 
Since the analytic form of $f_i(\boldsymbol{x})$ is unavailable, many common optimization techniques, such as Lagrange multiplier, do not apply.

We propose a solution to the optimization problem. 
Define
\begin{equation}
  g(\boldsymbol{x})=\sum_{i=0}^n R(f_i(\boldsymbol{x}))
  \label{eqn:joint}
\end{equation}
where the rectifier $R(x) \equiv 0 \vee x$ \ (the binary $\vee$ operator outputs the larger value of its operands).
Therefore, $g(\boldsymbol{x}) = 0$ only if $f_i(\boldsymbol{x}) = 0, \forall i\in [0, n]$. 
In other words, we combined the $n$ optimizations into one optimization. 
Now we can use the gradient descent algorithm, similar to the one used by Angora, to find a solution to $g(\boldsymbol{x}) = 0$. 
Note that when we compute the gradient of $g(\boldsymbol{x})$ using differentiation, we need to artificially keep the branch choices of the effective prior conditional statements of $s$ to ensure that $s$ is reachable.


Let us revisit the program in \autoref{fig:reachability}. Let $[x,y] = [1,3]$ be the initial input. When we fuzz the target conditional statement $s$ on Line~\ref{lst:main:br5} to explore the true branch, we cannot solve the branch constraint by mutating only $y$. Using joint optimization, we combine the branch constraints of $s$ and its effective prior conditional statements on Line~\ref{lst:main:br2} and \ref{lst:main:br1} to construct (by  \autoref{eqn:joint} and \autoref{tbl:trans}):
\[
  g([x, y]) = R(x-2+\epsilon) + R(x+y-3+\epsilon) + R(1-y+\epsilon)
\]
where $\epsilon=1$. On the initial input $[x,y] = [1,3]$, $g([x, y])=2$. Using gradient descent, we will find a solution to $g([x, y])=0$ where $[x,y] =[0,2]$.

\begin{table}[t]
\caption{Transform a predicate into a function such that the predicate is satisfied when the function is non-positive. $\epsilon$ is the smallest positive value of the type for $a$ and $b$. For integers, $\epsilon=1$.}
\begin{center}
\begin{tabular}{lll}
\toprule
Predicate & $f()$\\
\midrule
$a < b$ & $a-b+\epsilon$ \\
$a \le b$ & $a-b$ \\
$a > b$ & $b-a+\epsilon$ \\
$a \ge b$ & $b-a$ \\
$a = b$ & $\operatorname{abs}(a-b)$ \\
$a \not= b$ & $-\operatorname{abs}(a-b)+\epsilon$ \\
\bottomrule 
\end{tabular}
\end{center}
\label{tbl:trans}
\end{table}

\subsection{Detect implicit effective prior conditional statements}

\begin{figure}[t]
\begin{lstlisting}
void bar(int y, int z) {
    int k = 0, n = 0;
    if (z - y == 56789) { (*@\label{lst:imp:br2}@*)
        k = 1; n = 1;
    }
    if (k == 1) { (*@\label{lst:imp:br3}@*)
        if (z == 123456789) { .... }(*@\label{lst:imp:br4}@*)
    }
}

void foo(int x, int y, int z) {
    void (*fun_ptr)(int, int) = NULL; 
    if (z - x == 12345) { (*@\label{lst:imp:br1}@*)
        fun_ptr = &bar;
    } else {
        fun_ptr = &other_fn;
    }
    (*fun_prt)(y, z); (*@\label{lst:imp:indirectcall}@*)
}
\end{lstlisting}
\caption{A program showing implicit control and taint flow dependencies}
\label{fig:implicit_example}
\end{figure}

The mutation strategies in \autoref{sec:optimization} may fail if we cannot find all the control flow and taint flow dependencies among conditional statements. \autoref{sec:controlflow} and \autoref{sec:dataflow} described algorithms for finding all the \emph{explicit} control flow and taint flow dependencies, respectively. However, they are unable to find \emph{implicit} flows.  \autoref{fig:implicit_example} shows such an example. The conditional statement on
Line~\ref{lst:imp:br1} causes an implicit taint flow into \texttt{fun\_ptr} in function \texttt{foo},  which then implicitly determines the control flow whether the program will call the function \texttt{bar} or \texttt{other\_fn}. Also, Line~\ref{lst:imp:br2} causes an implicit taint flow into the variable \texttt{k}, whose value will determine the value of the predicate on Line \ref{lst:imp:br3}. Therefore, both the conditional statements on Line~\ref{lst:imp:br1} and Line~\ref{lst:imp:br2} should be effective prior conditional statements for the target statement on Line~\ref{lst:imp:br4}. However, since the taint flow is implicit, the algorithms in \autoref{sec:dataflow} cannot find them.

Implicit taint flows may be identified using control flow graphs~\cite{kang2011dta}. If a predicate is tainted, then the method taints all the variables that get new values in either branch of the conditional statement. For byte-level taint tracking, this method adds the taint label of the predicate to each of the above variables. For example, consider the predicate on Line~\ref{lst:imp:br2} in \autoref{fig:implicit_example}. Since the variables \texttt{k} and \texttt{n} are assigned new values in a branch of this conditional statement, this method adds the taint label of the predicate (i.e., the taint label of the variable \texttt{y}) to the variable \texttt{k} and \texttt{n}. However, this method often results in over taint or taint explosion, because it may add taint labels that will be useless to the analysis. For example, in the example above, while the taint label added to the variable \texttt{k} captures the implicit taint flow dependency from Line~\ref{lst:imp:br2} to Line~\ref{lst:imp:br3}, the taint label added to the variable \texttt{n} is useless because it does not help identify new taint flow dependencies between conditional statements. Even worse, these useless taint labels will propagate further to other parts of the program, resulting in taint explosion.

We propose a novel approach to identify implicit control flow and taint flow dependencies between conditional statements without incurring either huge analysis overhead or taint explosion. The insight is that rather than identifying all the implicit flows, we need to identify only those that cause the target conditional statement to become unreachable during input mutation. 
Let $s$ be the target conditional statement that was reachable on the original input but became unreachable on the mutated input. 
We run the program twice. First, we run the program on the original input and record the branch choices of all the conditional statements on the path before $s$. Then, we run the program on the mutated input with a special handling: when we encounter a conditional statement, we record its branch choice but force it to take the branch choice as in the previous run (on the original input). Therefore, the paths of the two runs have the same sequence of conditional statements. We examine all the conditional statements on the path from the start of the program to $s$ in the reverse chronological order. For each such statement $t$, if it is not already an explicit effective prior statement identified by the algorithms in \autoref{sec:dataflow} and if its branch choices in the first run (on the original input) and the second run (on the mutated input) differ, it has a potential control flow or taint flow dependency with $s$. To test whether this dependency truly exists, we run the program on the mutated input with a special handling: we force all the following conditional statements to take the branch choice as in the first run: 
\begin{enumerate}
  \item all the conditional statements before $t$ on the path
  \item all the explicit effective prior conditional statements
  \item all the implicit effective prior conditional statements
\end{enumerate}

If the program no longer reaches $s$, then $t$ truly has implicit control flow or taint flow dependency with $s$, and we mark it as an implicit effective prior conditional statement of $s$.

The complexity of this algorithm is linear in the number of conditional statements before $s$ that are affected by the mutated bytes but are not control-flow-wise dependent on $s$. However, since \name mutates inputs by gradient descent on a small proportion of the input, the number of statements we should test is likely to be few.

\section{Implementation}

We implemented \name in 8672 lines of Rust, and 1262 lines of C++ for LLVM pass. We borrowed from Angora the code for byte-level taint tracking and for mutating the input by gradient descent~\cite{angora}. When computing intraprocedural post dominator trees, \name uses LLVM's function pass \texttt{PostDominatorTree\allowbreak Wrapper\allowbreak Pass}~\cite{llvm}. 

\section{Evaluation}
\label{sec:evaluation}

We evaluated \name in three parts. In the first part, we compared \name with nine other fuzzers on the LAVA-M data set~\cite{lava2016}. Next, we compared \name with three other most representative fuzzers on 13 open source programs. Finally, we evaluated how \name's ability to solve nested constraints contributes to its impressive performance.

We ran our experiments on a server with two Intel Xeon Gold 5118 processors and 256 GB memory running 64-bit Debian 10.\footnote{\name does not need that much amount of memory. We also successfully fuzzed all the programs on our laptop with only 8 GB memory.} Even though \name can fuzz a program using multiple cores simultaneously, we configured it to fuzz the programs using only one core during evaluation.
We ran each experiment five times and reported the average performance.

\subsection{Comparison on LAVA-M}
LAVA-M consists of four programs with a large number of injected but realistic looking bugs~\cite{lava2016}. It has been widely used for evaluating fuzzers. However, it is approaching the end of its shelf life as the state of the art fuzzers (Angora and REDQUEEN) were able to find almost all the injected bugs in LAVA-M. While LAVA-M cannot show that \name advances the state of the art, it can show whether \name is at the state of the art.

\autoref{tbl:lava_all} compares the number of bugs found by 10 fuzzers. \name and REDQUEEN are the best: they both found almost all the listed bugs in LAVA-M.\footnote{\name and REDQUEEN also found several unlisted bugs, which the LAVA-M authors injected but were unable to trigger. \autoref{tbl:lava_unlisted} shows the IDs of unlisted bugs.} 

\begin{table*}[t]
  \caption{Bugs found on the LAVA-M data set by different fuzzers} 
  \begin{center}
  \setlength{\tabcolsep}{0.11cm} 
  \begin{tabular}{lSSSSSSSSSSS}
  \toprule
  \multirow{2}{*}{Program} & {Listed} & \multicolumn{10}{c}{Bugs found by each fuzzer}\\ \cmidrule(lr){3-12}
               & {bugs} & {AFL} & {FUZZER}  & {SES} & {VUzzer}  & {Steelix} & {QSYM} & {NEUZZ} & {REDQUEEN}    & {Angora}  & {\textbf{\name}} \\
  \midrule
  \file{uniq}   & 28      & 9   & 7  & 0  & 27     & 7    & 28     & 29     & 29      & 29    & 29     \\
  \file{base64} & 44      & 0   & 7  & 9  & 17     & 43   & 44     & 48     & 48      & 48    & 48     \\
  \file{md5sum} & 57      & 0   & 2  & 0  & {Fail} & 28   & 57     & 60     & 57      & 57    & 57     \\
  \file{who}    & 2136    & 1   & 0  & 18 & 50     & 194  & 1238   & 1582   & 2462    & 1541  & 2432   \\
  \bottomrule
  \end{tabular}
  \end{center}
  \label{tbl:lava_all}
  \end{table*}

\subsection{Comparison on 13 open source programs}
\label{sec:opensource}

We compared \name with AFL, QSYM and Angora by line and branch coverage. We ran them on 13 open source programs shown in \autoref{tbl:program_detail}. We chose these programs because eight of them were used for evaluating Angora, and the rest were used frequently for evaluating other fuzzers. 

\begin{table}[t] 
  \caption{Programs used in evaluation in \autoref{sec:opensource}}
  \begin{center}
  \begin{tabular}{lllS}
    \toprule
  
  Program & Version & Argument & {Size (kB)} \\
  \midrule

  \file{djpeg(ijg)}        & v9c               &               & 859 \\
  \file{file}              & commit-6367a7c9b4 &               & 781 \\
  \file{jhead}             & 3.03              &               & 205 \\
  \file{mutool(mupdf)}     & commit-08657851b6 & \texttt{draw} & 39682 \\
  \file{nm(binutils)}      & commit-388a192d73 & \texttt{-C}   & 6659 \\
  \file{objdump(binutils)} & commit-388a192d73 & \texttt{-x}   & 9357 \\
  \file{readelf(binutils)} & commit-388a192d73 & \texttt{-a}   & 2119 \\
  \file{readpng(libpng)}   & commit-0a882b5787 &               & 1033 \\
  \file{size(binutils)}    & commit-388a192d73 &               & 6597 \\
  \file{tcpdump(libpacp)}  & commit-e9439e9b71 & \texttt{-nr}  & 6022 \\
  \file{tiff2ps(libtiff)}  & commit-a0e273fdca &               & 1517 \\
  \file{xmllint(libxml2)}  & commit-d3de757825 &               & 6862 \\
  \file{xmlwf(expat)}      & commit-9f5bfc8d0a &               & 785 \\
  \bottomrule
  \end{tabular}
  \end{center}
  \label{tbl:program_detail}
  \end{table}

\subsubsection{Program coverage and efficiency}

We compared line and branch coverage of AFL (1 Master + 1 Slave), Angora (+1 AFL Slave), QSYM (+ 1 AFL Slave) with and without optimistic solving, and \name (+1 AFL Slave). \autoref{tbl:real_app} shows the coverage after running AFL, Angora, QSYM, QSYM with optimistic solving disabled, and \name on two CPU cores for 24 hours (one core for AFL slave). \name outperformed AFL, QSYM, and Angora on all the programs, except on \file{xmlwf}, \file{mutool}, and \file{tiff2ps} where \name had similar performance with Angora. \name's advantage shines the most on \file{xmllint}, where \name increased line and branch coverage by 16.8\% and 21.8\%, respectively, over Angora, the fuzzer with the next highest coverage. 

\begin{table*}[t] 
\caption{Comparison of coverage between AFL, QSYM, Angora and \name}
\begin{center}
\begin{tabular}{lrrrrrrrrrr}
  \toprule
  \multirow{3}{*}{Program} & \multicolumn{5}{c}{Line coverage} & \multicolumn{5}{c}{Branch coverage} \\ \cmidrule(lr){2-6} \cmidrule(lr){7-11}
    & \multirow{2}{*}{AFL}  & \multicolumn{2}{c}{QSYM}  & \multirow{2}{*}{Angora} & \multirow{2}{*}{\bfseries \name} & \multirow{2}{*}{AFL} & \multicolumn{2}{c}{QSYM} & \multirow{2}{*}{Angora} & \multirow{2}{*}{\bfseries \name} \\ \cmidrule{3-4} \cmidrule{8-9}
& & opt on & opt off & & & & opt on & opt off \\
\midrule
    \file{djpeg}   & 5951  & 5994  & 5967   & 5900  & 6144  & 1915  & 1899 & 1910  & 1855  & 2123 \\ 
    \file{file}    & 2637  & 3098  & 2799   & 3179  & 3277  & 1746  & 2073 & 1887  & 2102  & 2284 \\
    \file{jhead}   & 399   & 761   & 756    & 903   & 948   & 218   & 445  & 440   & 538   & 571  \\ 
    \file{mutool}  & 5247  & 5557  & 5493   & 5631  & 5694  & 2177  & 2429 & 2366  & 2495  & 2550 \\
    \file{nm}      & 4766  & 6002  & 5390   & 6261  & 6964  & 2765  & 3314 & 3122  & 3452  & 3866 \\
    \file{objdump} & 3904  & 6380  & 5678   & 7906  & 8076  & 2291  & 3190 & 3090  & 4263  & 4297 \\
    \file{readelf} & 7792  & 8357  & 7906   & 10203 & 11245 & 5810  & 5645 & 5863  & 7243  & 7993 \\
    \file{readpng} & 1643  & 2047  & 1723   & 2027  & 2187  & 903   & 1142 & 956   & 1161  & 1278 \\
    \file{size}    & 3299  & 4960  & 3845   & 5332  & 5445  & 1937  & 2438 & 2222  & 2893  & 2965 \\
    \file{tcpdump} & 13000 & 12485 & 13362  & 13691 & 13992 & 7455  & 7025 & 7630  & 8004  & 8210 \\ 
    \file{tiff2ps} & 5193  & 4892  & 5054   & 5303  & 5291  & 3217  & 3048 & 3109  & 3325  & 3304 \\ 
    \file{xmllint} & 5804  & 6221  & 6058   & 6516  & 7611  & 4877  & 5334 & 5129  & 5786  & 7045 \\
    \file{xmlwf}   & 4850  & 4732  & 4684   & 5011  & 5019  & 1965  & 1920 & 1886  & 2042  & 2041 \\ 
\bottomrule
\end{tabular}
\end{center}
\label{tbl:real_app}
\end{table*}

\begin{figure*}[t]
\centering
\begin{subfigure}[b]{0.48\linewidth}
  \includegraphics[width=\textwidth]{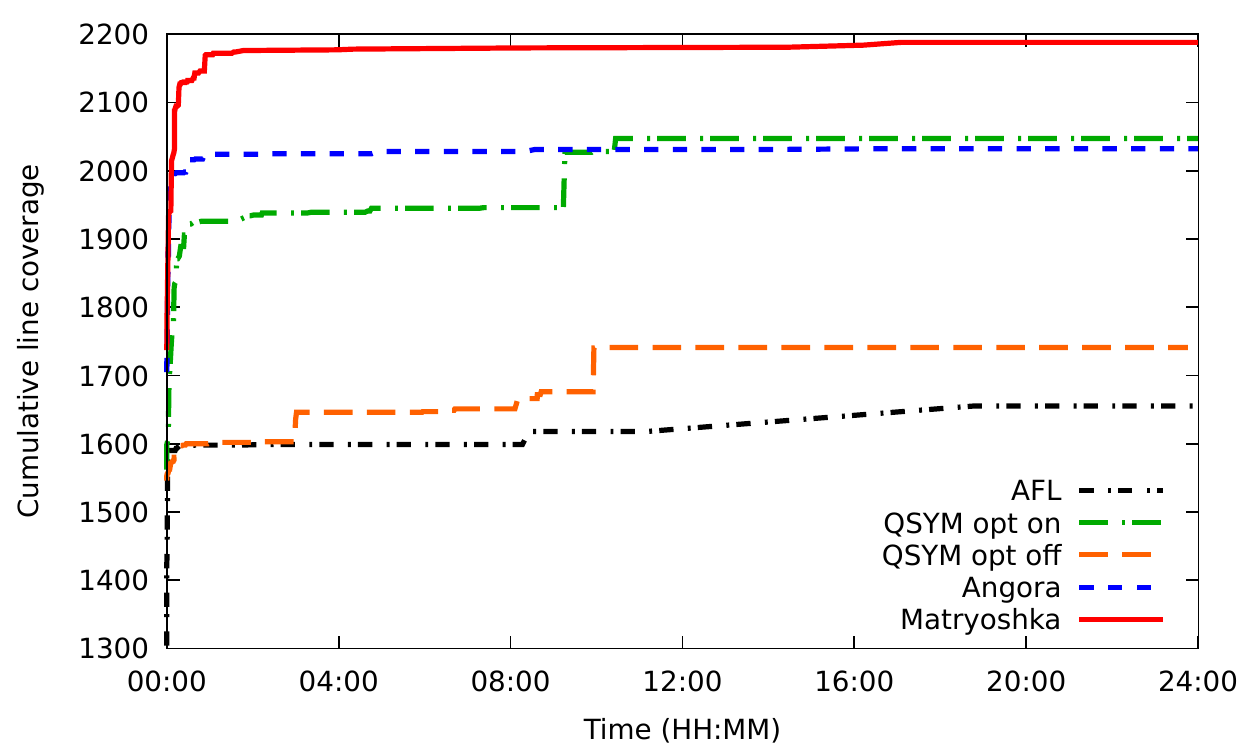}
  \caption{Line coverage}
\end{subfigure}
\begin{subfigure}[b]{0.48\linewidth}
  \includegraphics[width=\textwidth]{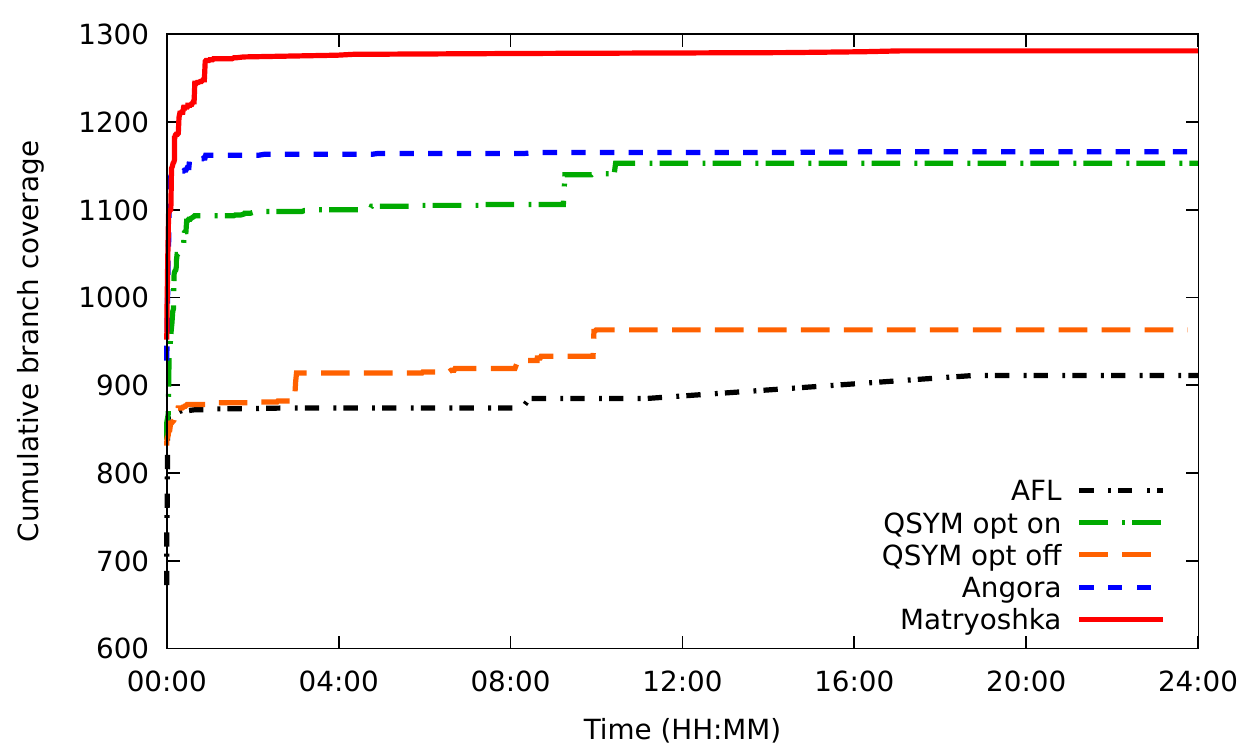}
  \caption{Branch coverage}
\end{subfigure}
\caption{Cumulative line and branch coverage on \file{readpng} by AFL, QSYM, Angora, and \name in 24 hours}
\label{fig:readpng_cov}
\end{figure*}

\autoref{fig:readpng_cov} compares the cumulative line and branch coverage by AFL, Angora, and \name on the program \file{readpng} over time. 
\name covered more lines and branches than QSYM and Angora at all time, thanks to its ability to solve nested conditional statements.

The goal of coverage-based fuzzers is to increase coverage, as measured by cumulative line and branch coverage. By contrast, the number of tests generated and executed by the fuzzer per second has no correlation with either line or branch coverage across different fuzzers, because smart fuzzers may generate fewer tests but the tests are much more effective in triggering new branches.

\subsubsection{Bug analysis, verification, and classification}

Besides all the inputs that crashed the program during fuzzing, we also ran AddressSanitizer(ASAN)~\cite{asan} on all the seeds found by \name and saved the inputs where ASAN reported errors. Then, we deduplicated the crashes by AFL's \texttt{afl-cmin -C} command.

We manually verified all the crashes and classified them into unique bugs shown in \autoref{tbl:bug_analysis}. \name found a total of 41 unique bugs in seven programs (it found no bugs in the rest six programs). 
We have reported all the bugs to the developers and 12 of them have been assigned CVE IDs.

\begin{table}[t] 
  \caption{Classification of verified bugs found by \name. SBO: stack buffer overflow; HBO: heap buffer overflow; OOM: out of memory; OBR: out of bound read.}
  \begin{center}
    \begin{tabular}{lccccc}
      \toprule
      \multirow{2}{*}{Program}  & \multicolumn{5}{c}{Number of bugs}\\
                                & SBO & HBO & OOM & OBR & Total     \\
      \midrule
      \file{file}    & 4 &    &   &   & 4  \\
      \file{jhead}   & 2 & 15 &   & 6 & 23 \\
      \file{nm}      & 1 & 1  &   &   & 2  \\
      \file{objdump} &   &    & 3 & 1 & 4  \\
      \file{size}    &   & 1  & 1 &   & 2  \\
      \file{readelf} &   & 4  &   &   & 4  \\
      \file{tiff2ps} &   & 1  & 1 &   & 2  \\
      \bottomrule
    \end{tabular}
  \end{center}
  \label{tbl:bug_analysis}
\end{table}

\subsection{Novel features of \name}

We evaluated the key novel feature of \name: its ability to solve constraints involving nested conditional statements.

\subsubsection{Solved constraints}
\label{sec:solvedconstaints}

On each program and given the same seeds, the constraints that \name can solve is a superset of the constraints that Angora can solve. This is because for each constraint, \name will first try to solve it using Angora's method. If it fails, then \name will start to use the methods in \autoref{sec:design}. We evaluated which constraints unsolved by Angora could be solved by \name. To eliminate the impact of randomness on the fuzzers, we collected the inputs generated by AFL and fed them as the only seeds to both Angora and \name. In other words, we discarded the new seeds generated by Angora and \name during fuzzing, respectively. We ran \name using three different mutation strategies described in \autoref{sec:optimization} for five hours: \emph{prioritize reachability} (\autoref{sec:prioritizereachability}), \emph{prioritize satisfiability} (\autoref{sec:prioritizesatisfiability}), and \emph{joint optimization} (\autoref{sec:joint}). 

\autoref{tbl:solve_ability_new_vs_old} shows the number of constraints that \name could solve but Angora could not. The table shows that \name could solve as few as 172 and as many as 1794 new constraints (that were unsolvable by Angora) per program. This demonstrates the effectiveness of the algorithms in \autoref{sec:design}. \autoref{tbl:solve_ability_strategies} compares \name's three strategies for solving constraints described in \autoref{sec:optimization}. The strategy prioritizing satisfiabily was the most effective, but there were constraints that this strategy could not solve but others could. The strategy prioritizing reachability was effective on \file{jhead} and \file{size}, and the joint optimization strategy was effective on \file{readpng}. 

\begin{figure}[t]
  \centering
    \includegraphics[width=0.48\textwidth]{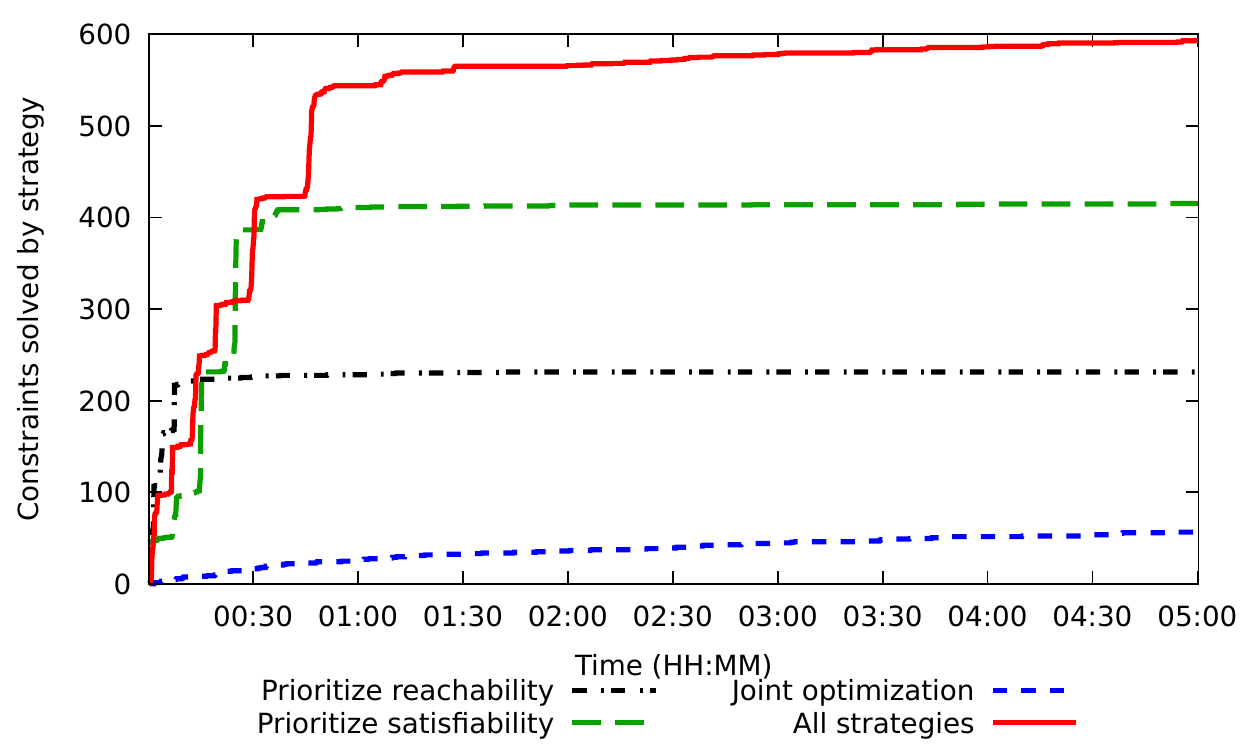}
    \caption{Cumulative constraints solved by \name's three strategies, respectively, in five hours on the program \file{size}. \emph{All strategies} means trying prioritizing reachability, prioritizing satisﬁability, and joint optimization, in that order until the constraint is solved.}
  \label{fig:solved_constraints}
  \end{figure}

\autoref{fig:solved_constraints} compares the cumulative constraints solved by each individual strategy over five hours on the program \file{size}. We can see that the strategies prioritize reachability (PR) and prioritize satisfiability (PS) contribute greatly to the the number of constraints solved early on in fuzzing, while joint optimization (JO) solves constraints slowly but continues to grow later on when the other two strategies have reached their respective plateaus.

\begin{table}[t] 
  \caption{Constraints unsolved by Angora, and nested constraints unsolved by Angora but solved by \name}
\begin{center}
\begin{tabular}{lSSSSSS}
  \toprule

    \multirow{3}{*}{Program} & \multicolumn{2}{c}{Unsolved by} & {Solved by} & {\% of nested} \\
     & \multicolumn{2}{c}{Angora} & {\name} & {constraints}\\
     \cmidrule(lr){2-3} \cmidrule(lr){4-4} & {All} & {Nested} & {Nested} & {solved} \\
\midrule
    \file{djpeg}   & 1889 & 1700 & 345  & \SI{20.3}{\percent} \\ 
    \file{file}    & 610  & 527  & 172  & \SI{32.6}{\percent} \\ 
    \file{jhead}   & 4923 & 2853 & 316  & \SI{11.1}{\percent} \\ 
    \file{mutool}  & 1883 & 1523 & 249  & \SI{16.3}{\percent} \\ 
    \file{nm}      & 2564 & 2162 & 408  & \SI{18.9}{\percent} \\ 
    \file{objdump} & 4418 & 4000 & 377  & \SI{9.4}{\percent} \\ 
    \file{readelf} & 4012 & 3375 & 621  & \SI{18.4}{\percent} \\ 
    \file{readpng} & 5353 & 5033 & 1170 & \SI{23.2}{\percent} \\ 
    \file{size}    & 4359 & 3830 & 593  & \SI{15.5}{\percent} \\ 
    \file{tcpdump} & 4343 & 4079 & 1794 & \SI{44.0}{\percent} \\ 
    \file{tiff2ps} & 8923 & 6564 & 330  & \SI{5.0}{\percent} \\ 
    \file{xmllint} & 1838 & 1437 & 271  & \SI{18.9}{\percent} \\ 
    \file{xmlwf}   & 5233 & 5033 & 301  & \SI{6.0}{\percent} \\ 
\bottomrule
\end{tabular}
\end{center}
\label{tbl:solve_ability_new_vs_old}
\end{table}

\begin{table}[t] 
    \caption{Constraints solved by \emph{prioritizing reachability} (PR, \autoref{sec:prioritizereachability}), \emph{prioritizing satisfiability} (PS, \autoref{sec:prioritizesatisfiability}), and \emph{joint optimization} (JO, \autoref{sec:joint}).}
\begin{center}
\begin{tabular}{lSSSSSS}
  \toprule
  \multirow{2}{*}{Program} &
  \multicolumn{3}{c}{Constraints solved by}\\ 
&  {PR} &  {PS}  & {JO} \\
\midrule
\file{djpeg}    & 1   & 305  & 72    \\ 
\file{file}     & 5   & 163  & 11    \\ 
\file{jhead}    & 172 & 243  & 60    \\ 
\file{mutool}   & 1   & 247  & 12   \\ 
\file{nm}       & 30  & 321  & 78    \\ 
\file{objdump}  & 47  & 343  & 53    \\ 
\file{readelf}  & 2   & 573  & 86    \\ 
\file{readpng}  & 0   & 1043 & 313   \\ 
\file{size}     & 231 & 414  & 56    \\ 
\file{tcpdump}  & 20  & 1742 & 59    \\ 
\file{tiff2ps}  & 10  & 323  & 16   \\ 
\file{xmllint}  & 1   & 252  & 31   \\ 
\file{xmlwf}    & 1   & 253  & 97    \\ 
\bottomrule
\end{tabular}
\end{center}
\label{tbl:solve_ability_strategies}
\end{table}

\autoref{fig:crc_example} shows an example where Angora could not reach the false branch of Line~\ref{lst:main:value_check} but \name could. This is because when Angora mutated \texttt{buffer[0]} to satisfy the false branch of 
Line~\ref{lst:main:value_check}, it caused the CRC check on Line~\ref{lst:main:crc_check} to fail, so the function never reached Line~\ref{lst:main:value_check}. Using the strategy for prioritizing satisfiability, \name first found an assignment to \texttt{buffer[0]}, either 1 or 2, to reach the false branch of Line~\ref{lst:main:value_check}. Then, it backtracked to the prior conditional statement on Line~\ref{lst:main:crc_check}. Through byte-level taint analysis, \name learned that all the input bytes flowed into Line~\ref{lst:main:crc_check}, but since \texttt{buffer[0]} also flowed into Line~\ref{lst:main:value_check}, this strategy directed \name to keep \texttt{buffer[0]} fixed but to freely mutate all the bytes. Using gradient descent, \name found an input that satisfied the false branch of Line~\ref{lst:main:crc_check}.\footnote{The reason why gradient descent helped \name to find a solution on Line~\ref{lst:main:crc_check} is that the CRC value itself was in the input. Therefore, gradient descent found that the objective function had a constant partial derivative with regard to the input bytes containing the CRC value, so it directed \name to mutate those input bytes to reduce the objective function to zero.}

\subsubsection{Effective prior conditional statements}

A key insight that allows \name to solve nested constraints effectively is that it identifies effective prior conditional statements, whose branch choices may cause the target conditional statement to become unreachable, and solves a constraint that consists of only those statements, instead of all the prior conditional statements on the path as done in traditional symbolic execution.  \autoref{tbl:nested_cond} compares the average number of effective prior conditional statements vs all prior conditional statements. It shows that the effective prior conditional statements account for a very small fraction of all the prior conditional statements (less than 5\% on 11 programs, and less than 10\% on all the 13 programs). This fact significantly reduces the complexity of the path constraints that \name solves and increases the likehood that the constraints can be solved.

\begin{table}[t] 
  \caption{Number of average effective prior conditional statements vs.\ all prior conditional statements}
  \begin{center}
  \begin{tabular}{lSSS}
    \toprule
    \multirow{2}{*}{Program} &  \multicolumn{3}{c}{Average prior conditional statements} \\
    \cmidrule{2-4}
    & {Effective}  & {All} & {Effective/all}\\
    
  \midrule
  \file{djpeg}   & 21.69 & 1217.98 &\SI{1.8}{\percent} \\ 
  \file{file}    & 22.27 & 345.25 &\SI{6.5}{\percent} \\ 
  \file{jhead}   & 16.81 & 2425.00 &\SI{0.7}{\percent} \\ 
  \file{mutool}  & 20.08 & 2087.80 &\SI{1.0}{\percent} \\ 
  \file{nm}      & 27.93 & 842.54  &\SI{3.3}{\percent} \\ 
  \file{objdump} & 23.93 & 493.24  &\SI{4.9}{\percent} \\ 
  \file{readelf} & 7.23  & 2498.21 &\SI{0.3}{\percent} \\ 
  \file{readpng} & 21.18 & 859.02  &\SI{2.5}{\percent} \\ 
  \file{size}    & 21.72 & 469.46  &\SI{4.6}{\percent} \\ 
  \file{tcpdump} & 26.26 & 268.52  &\SI{9.8}{\percent} \\ 
  \file{tiff2ps} & 30.44 & 1747.16 &\SI{1.7}{\percent} \\ 
  \file{xmllint} & 11.80 & 502.39  &\SI{2.3}{\percent} \\ 
  \file{xmlwf}   & 5.88  & 655.31  &\SI{0.9}{\percent} \\ 
  \bottomrule
  \end{tabular}
  \end{center}
  \label{tbl:nested_cond}
  \end{table}

\section{Discussion}

\subsection{Comparison with concolic execution}

We compare \name with QSYM while its last branch solving is disabled. This directly compares the effectiveness of \name's optimization strategies to that of a concolic execution engine. \autoref{tbl:real_app} shows that \name performs better than QSYM in all the statistics. This demonstrates that prioritizing reachability, satisfiability, and joint optimization can be used on most path constraints effectively without having to resort to concolic execution.

\subsection{Unsolved constraints}
\label{sec:unresolved_constaints}

\subsubsection{Unsatisfiable constraints}

Some constraints are unsatisfiable. \autoref{fig:conflict_example} shows an example in \file{readpng}. The program calls \texttt{png\_check\allowbreak \_chunk\_name} before calling \texttt{png\_format\_buffer}. \texttt{png\_check\allowbreak \_chunk\allowbreak \_name} checks if the character is alphanumerical on Line~\ref{lst:check1}. If not, it exits with an error. But later \texttt{png\_format\_buffer} checks the character again on on Line~\ref{lst:check2}, so the false branch of this line is unsatisfiable.

\begin{figure}[t]
\begin{lstlisting}[xleftmargin=0.5em]
// pngrutilc.c
void 
png_check_chunk_name(png_const_structrp png_ptr, 
            const png_uint_32 chunk_name) {
  for (i=1; i<=4; ++i) {
    if (c < 65 || c > 122 || 
        (c > 90 && c < 97)) (*@\label{lst:check1}@*)
      png_chunk_error(png_ptr, 
        "invalid chunk type");
    ...
  }
  ..
}
// pngerror.c 445
void 
png_format_buffer(png_const_structrp png_ptr, 
            png_charp buffer,
            png_const_charp error_message) {
  ...
  if (isnonalpha(c) != 0) { ... }(*@\label{lst:check2}@*)
}
\end{lstlisting}
\caption{An example with an unsatisfiable constraint. The false branch on Line~\ref{lst:check2} is unsatisfiable because it is precluded by an earlier check on Line~\ref{lst:check1}. }
\label{fig:conflict_example}
\end{figure}

\subsubsection{Taint lost in propagation}

\label{sec:taint_lost}

\autoref{sec:dataflow} uses the results from byte-level taint tracking to determine the taint flow dependency between nested conditional statements. Similar to Angora, \name also extended DFSan~\cite{dfsan} to implement byte-level taint tracking, but neither of the two is able to track taint flows through external libraries. We manually modeled the taint flow in common external libraries for \name, but this is in no way comprehensive.

\subsubsection{Program crashing when applying the strategy for prioritizing satisfiability and joint optimization}

When mutating the input using the strategy for prioritizing satisfiability (\autoref{sec:prioritizesatisfiability}) and joint optimization (\autoref{sec:joint}), \name artificially keeps the branch choices of prior conditional statements. This may cause the program to crash. For example, a conditional statement may serve to prevent the program from accessing data out of bound. If we mutate the length of the data but artificially keep the branch choice of the conditional statement, the program may access data out of bound and crash.

\subsubsection{Difficult joint constraints}

The joint optimization strategy is the last resort for mutation. We examined the conditional statements that have at least one effective prior conditional statement in the program \file{tiff2ps} and found that they have on average 30 such prior statements in \autoref{tbl:nested_cond}. It is difficult to solve such a complex joint constraint.

\subsubsection{Constraint dependent on order of branches}

\label{sec:branch_order}

On \file{xmlwf}, \name and Angora reached similar branches. The unreached bra\-nches are guarded by predicates that can only be solved through a specific combination of other branch choices, a situation that none of the fuzzers we tested are designed to handle. These situations are commonly seen in parser logic, where a conditional statement checks the internal state of the parser, while the current state depends on the order of the branches reached.

\subsection{Other limitations of \name}

\subsubsection{Design limitations}

\name's branch counting method is derived from AFL's, a coarse grained method that can only provide limited information about the program's internal state. This is to maintain compatibility with AFL and AFL-like fuzzers for synchronization, but leads to issues such as those mentioned in \autoref{sec:branch_order}.

\subsubsection{Implementation limitations}

The current implementation of \name requires source code because we use compile-time instrumentation. We could overcome this limitation by instrumenting the executables. \name's taint tracking uses byte-level granularity as a balance between efficiency and accuracy, as bit-level taint tracking would require significantly more memory and computing power. \autoref{sec:unresolved_constaints} described other implementation limitations.

\section{Related work}

\subsection{Solving complicated constraints}

Symbolic execution has the potential to solve complex constraints~\cite{cadar2008klee, cha2012unleashing} and is used in fuzzing~\cite{godefroid2005dart, godefroid2008automated, cha2015program_adaptive, li2017steelix, redqueen, stephens2016driller, tfuzz, wang2010taintscope, qsym}. One example is Driller, which uses symbolic execution only when the co-running AFL cannot progress due to complicated constrains~\cite{stephens2016driller}. Steelix~\cite{li2017steelix} and REDQUEEN~\cite{redqueen} detect magic bytes checking and infer their input offsets to solve them without taint analysis. T-Fuzz ignores input checks in the original program and leverages symbolic execution to filter false positives and reproduce true bugs~\cite{tfuzz}. TaintScopre fixs checksum values in the generated inputs using symbolic execution~\cite{wang2010taintscope}. In T-Fuzz and TaintScope, input checks and checksum checks are complex constraints. However, symbolic execution faces the challenges of path explosion and scalability~\cite{cadar2013symbolic, shoshitaishvili2016sok}. QSYM uses fast concolic execution to overcome the scalability problem, but similar to Angora, it solves only the constraint of the target conditional statement without considering any nesting relationships between other conditional statements~\cite{qsym}.
By contrast, \name finds all those nesting conditional statements and searches for an input that satisfies all of them. 

\subsection{Using control flow to guide fuzzing}

Run-time control flow can contain information useful for guiding fuzzing~\cite{vuzzer2017, tfuzz, wang2010taintscope, angora2018, fairfuzz, aflgo, hawkeye, youprofuzzer, gan2018collafl}. VUzzer uses control flow information to prioritize inputs that may explore deep code blocks but that do not lead to error handling codes~\cite{vuzzer2017}. Angora prioritizes fuzzing on unexplored branches~\cite{angora2018}. AFLGo and Hawkeye measure the distance between the seed input and the target location in the control flow graph, and minimizes the distance in fuzzing~\cite{aflgo, hawkeye}. T-Fuzz~\cite{tfuzz} and TaintScope~\cite{wang2010taintscope} use control flow features to find sanity checks and checksum checks, respectively. After that, they remove these checks to cover more code. 

FairFuzz identifies the ``rare branches'' exercised by few inputs using control flow information and schedules the fuzzer to generate inputs targeting the ``rare branches''~\cite{fairfuzz}.
If a path constraint does not exhibit taint flow dependencies on the ``rare branches'', FairFuzz can solve them efficiently similar to QSYM and Angora. Otherwise, the input bytes flowing into the path constraint will not be included in the mutation mask, e.g. nested conditional statements, and FairFuzz will experience difficulties while solving it.

Post dominator trees~\cite{allen1970control} were used to determine control flow dependencies~\cite{ferrante1987program}. 
Xin et al.\cite{xin2007efficient} proposed a method to capture both intraprocedural and interprocedural control dependencies efficiently based on post-dominator trees. The method inserts code at the point before each conditional statement(\fn{branching}) and the head of its immediate post dominator block(\fn{merging}). Similarly, \name proposes an equivalent approach without the injections at the \fn{merging}, which is more efficient in our case of finding all the prior conditional statements.

SYMFUZZ~\cite{cha2015program_adaptive} uses control dependencies to infer input bit dependencies and use it to find an optimal mutation ratio for fuzzing. Under this method, nested conditional statements will introduce more complex input bit dependencies. SYMFUZZ utilizes this information to reduce mutation ratio for fuzzing, which is incapable of solving nested conditional statements efficiently.

\subsection{Using taint tracking to guide fuzzing}

Taint tracking can locate which input bytes flow into the security-sensitive code that may trigger bugs~\cite{ganesh2009taint, bekrar2012taint, haller2013dowsing}. VUzzer~\cite{vuzzer2017} is an application-aware fuzzer that uses taint analysis to locate the position of ``magic bytes'' in input files and assigns these ``magic bytes'' to fixed positions in the input. VUzzer can find ``magic bytes'' only when they appear continuously in the input. TIFF~\cite{jain2018tiff} is an improvement over VUzzer, using in-memory data structure identification techniques and taint analysis to infer input types. Angora~\cite{angora2018} tracks the flow of input bytes into conditional statements and mutates only those bytes. \name uses the same technique to identify relevant bytes in the input. Taintscope~\cite{wang2010taintscope} uses taint tracking to infer checksum-handling code and bypasses these chec\-ks using control flow alteration since these checks are hard to satisfy when mutating the input. T-Fuzz~\cite{tfuzz} detects complex checks without taint tracking. Both approaches use symbolic execution to generate valid input that would solve target constraints. Checksum-handling code is an classic example of nesting conditional statements: the code that uses the value is nested under the conditional statement that verifies the checksum. \name is able to handle such code naturally.

DTA++ allows dynamic taint analysis to avoid under-tainting when implicit flows occur in data transformations\cite{kang2011dta}. It locates culprit implicit flows that cause the under-tainting through binary search and generates rules to add additional taint for those control dependencies only.
However, the method may result in over-tainting. \name proposes an efficient approach that can avoid over-tainting when determining taint flow dependencies among conditional statements.

\subsection{Using machine learning to guide fuzzing}

Both Angora and \name view solving constraints as a search problem and take advantage of commonly used search algorithms in machine learning. Skyfire~\cite{wang2017skyfire} learns a probabilistic context-sensitive grammar (PCSG) from existing samples and leverages the learned grammar to generate seed inputs. Learn \& Fuzz~\cite{godefroid2017learn} first attempts to use a neural network to automatically generate an input grammar from sample inputs. Instead of learning a grammar, Rajal et al.\ use neural networks to learn a function to predict the promising input bytes in a seed file to perform mutations~\cite{rajpal2017not}. Konstantin et al.\ formalizes fuzzing as a reinforcement learning problem using the concept of Markov decision processes and constructs an algorithm based on deep $Q$-learning that chooses high reward actions given an input seed~\cite{bottinger2018deep}. 
NEUZZ~\cite{neuzz} uses a surrogate neural network to smoothly approximate a target program’s branch behavior and then generates new input by gradient-guided techniques to uncover new branches.

\subsection{Fuzzing without valid seed inputs}

SLF~\cite{You:2019:SFW:3339505.3339597} fuzzes programs without requiring valid inputs. It groups input bytes into fields where a field consists of consecutive bytes that affect the same set of checks. Then, it correlates checks whose predicates are affected by the same field. Finally, it uses a gradient-based method to mutate the fields to satisfy all the correlated checks. At a high level, SLF’s approach is comparable to \name s strategy of prioritizing satisfiability (\autoref{sec:prioritizesatisfiability}). The differences between \name and SLF are as follows. First, \name uses dynamic taint tracking to determine the bytes that flow into a predicate, while SLF uses probing. During probing, the SLF must flip each input byte individually, so if the input has $n$ bytes, then the program must run $n$ times. In contrast, dynamic taint tracking runs the program only once. Second, SLF determines the correlation between two checks based on their common input fields. However, this ignores their control flow flow dependency and may find unnecessary correlations. In contrast, \autoref{sec:controlflow} describes how \name determines the prior checks that the current check depends on by control flow. Third, SLF classifies some common checks into several categories and applies category-specific strategies effectively. For example, SLF can test offset/count of certain fields. By contrast, \name needs no prior knowledge of the types of checks and handles all checks uniformly. Finally, besides the strategy of prioritizing satisfiability, which is comparable to SLF’s strategy, \name also provides the strategies of prioritizing reachability and of joint optimization. \autoref{tbl:solve_ability_strategies} shows that these three strategies are complementary: together they can solve many more constraints than any single one of them can.

\section{Conclusion}

Deeply nested branches present a great challenge to coverage-based fuzzers. We designed and implemented \name, a tool for fuzzing deeply nested conditional statements. We proposed algorithms for identifying nesting conditional statements that the target branch depends on by control flow and taint flow, and proposed three strategies for mutating the input to solve path constraints. Our evaluation shows that \name solved more constraints and increased line and branch coverage significantly. \name found 41 unique new bugs in 13 open source programs and obtained 12 CVEs.

\section{Acknowledgment}

We thank Dongyu Meng for helpful discussions.

This material is based upon work supported by the National Science Foundation under Grant No.\ 1801751.

This research was partially sponsored by the Combat Capabilities Development Command Army Research Laboratory and was accomplished under Cooperative Agreement Number W911NF-13-2-0045 (ARL Cyber Security CRA). The views and conclusions contained in this document are those of the authors and should not be interpreted as representing the official policies, either expressed or implied, of the Combat Capabilities Development Command Army Research Laboratory or the U.S. Government. The U.S. Government is authorized to reproduce and distribute reprints for Government purposes not withstanding any copyright notation here on.

\setlength{\emergencystretch}{8em}

\printbibliography
\appendix
\onecolumn
\section*{Appendix}

\begin{table*}[hbt] 
  \caption{IDs of bugs injected but unlisted by LAVA, because the LAVA authors were unable to trigger them when preparing the data set. \name found these bugs.} 
  \begin{center}
  \begin{tabular}{lp{0.75\linewidth}}
  \toprule
  Program & IDs of bugs unlisted by LAVA-M but found by \name \\
  \midrule
  \file{uniq} & 227\\
  \file{base64} & 274, 521, 526, 527 \\
  \file{md5sum} & - \\
  \file{who} & 2, 4, 6, 8, 12, 16, 20, 24, 55, 57, 59, 61, 63, 73, 77, 81, 85, 89, 117, 125, 165, 169, 173, 177, 181, 185, 189, 193, 197, 210, 214, 218, 222, 226, 294, 298, 303, 307, 312, 316, 321, 325, 327, 334, 336, 338, 346, 350, 355, 359, 450, 454, 459, 463, 468, 472, 477, 481, 483, 488, 492, 497, 501, 504, 506, 512, 514, 522, 526, 531, 535, 974, 975, 994, 995, 996, 1007, 1026, 1034, 1038, 1049, 1054, 1071, 1072, 1329, 1334, 1339, 1345, 1350, 1355, 1361, 1377, 1382, 1388, 1393, 1397, 1403, 1408, 1415, 1420, 1429, 1436, 1445, 1450, 1456, 1461, 1718, 1727, 1728, 1735, 1736, 1737, 1738, 1747, 1748, 1755, 1756, 1891, 1892, 1893, 1894, 1903, 1904, 1911, 1912, 1921, 1925, 1935, 1936, 1943, 1944, 1949, 1953, 1993, 1995, 1996, 2000, 2004, 2008, 2012, 2014, 2019, 2023, 2027, 2031, 2034, 2035, 2039, 2043, 2047, 2051, 2055, 2061, 2065, 2069, 2073, 2077, 2079, 2081, 2083, 2085, 2147, 2181, 2189, 2194, 2198, 2219, 2221, 2222, 2223, 2225, 2229, 2231, 2235, 2236, 2240, 2244, 2246, 2247, 2249, 2253, 2255, 2258, 2262, 2266, 2268, 2269, 2271, 2275, 2282, 2286, 2291, 2295, 2302, 2304, 2462, 2463, 2464, 2465, 2466, 2467, 2468, 2469, 2499, 2500, 2507, 2508, 2521, 2522, 2529, 2681, 2682, 2703, 2704, 2723, 2724, 2742, 2790, 2796, 2804, 2806, 2810, 2814, 2818, 2823, 2827, 2834, 2838, 2843, 2847, 2854, 2856, 2915, 2916, 2917, 2918, 2919, 2920, 2921, 2922, 2974, 2975, 2982, 2983, 2994, 2995, 3002, 3003, 3013, 3021, 3082, 3083, 3099, 3185, 3186, 3187, 3188, 3189, 3190, 3191, 3192, 3198, 3202, 3209, 3213, 3218, 3222, 3232, 3233, 3235, 3237, 3238, 3239, 3242, 3245, 3247, 3249, 3252, 3256, 3257, 3260, 3264, 3265, 3267, 3269, 3389, 3439, 3443, 3464, 3465, 3466, 3467, 3468, 3469, 3470, 3471, 3487, 3488, 3495, 3496, 3509, 3510, 3517, 3518, 3523, 3527, 3545, 3551, 3561, 3939, 4224, 4287, 4295\\
  \bottomrule
  \end{tabular}
  \end{center}
  \label{tbl:lava_unlisted}
  \end{table*}
\end{document}